\documentclass{revtex4}
\usepackage{url}
\usepackage{xcolor}
\usepackage{hyperref}
\colorlet{shadecolor}{gray!15}
\definecolor{greenLinks}{rgb}{0,0.6,0}
\definecolor{blueLinks}{rgb}{0,0,0.6}
\definecolor{redLinks}{rgb}{0.6,0,0}
\definecolor{tempText}{rgb}{0.55,0.10,0.67}
\definecolor{eprintLinks}{rgb}{0.4,0.4,0.4}
\definecolor{journalLinks}{rgb}{0.6,0,0}
\usepackage{amssymb}
\usepackage{amsmath,color}
\usepackage{epsfig}
\usepackage{graphicx,epsf,epsfig}
\usepackage{footnote}
\usepackage{multirow}
\usepackage{graphicx}
\usepackage{subfigure}
\usepackage{lipsum}
\usepackage{multirow} 
\usepackage{tabulary}
\usepackage{rotating}
\usepackage{array}
\usepackage{color}

\def\det{\textrm{Det}}
\def\diag{\textrm{diag}}
\def\vpmns{{\mathbf{V}}_{\rm PMNS}}

\begin{document}

% Title portion
\title{Analysis of the Lepton Mixing Matrix in the Two Higgs Doublet Model}
%HDM-III}
%

%%%%%%%%%%%%%%%%%%%%%%%%%%%%%%%%%%%%%%%%%%%%%%%%%%%%%%%%%%%%%%%%%%%%%%%%%%%%%%%%%%%%%
%\author[aff1]{\bf E. Barradas-Guevara}\corref{cor1}
%%\eaddress[url]{barradas@fcfm.buap.mx}
%\author[aff2]{O. F\'elix-Beltr\'an}
%\eaddress{olga.felix@correo.buap.mx}
%%\eaddress{olga.felix@correo.buap.mx}
%\author[aff3]{ F. Gonzalez-Canales}
%\eaddress{fgonzalez@fis.cinvestav.mx}
%\author[aff1]{ E. Gonz\'alez-Hern\'andez}
%%\eaddress{felix.gonzalez@ific.uv.es}}
%\author[aff4]{ E. Rodr\'{\i}guez-J\'auregui}
%%\eaddress{olga.felix@correo.buap.mx}}
%\author[aff1]{M. Zeleny-Mora}
%%\eaddress{olga.felix@correo.buap.mx}}

%\affil[aff1]{Fac. de Cs. F\'{\i}sico Matem\'aticas, 
% Benem\'erita Universidad Aut\'onoma de Puebla,
% Apdo. Postal 1152, \\ Puebla, Pue.  72000, M\'exico.}
%\affil[aff2]{Fac. de Cs. de la Electr\'onica, 
% Benem\'erita Universidad Aut\'onoma de Puebla, 
% Apdo. Postal 542,\\ Puebla, Pue. 72000, M\'exico.}
%\affil[aff3]{Departamento de F\'{\i}sica, Centro de 
% Investigaci\'on y de Estudios Avanzados del Instituto Polit\'ecnico 
% Nacional, Apartado Postal 14-740, 07000 M\'exico D.F., M\'exico}
%\affil[aff4]{Departamento de F\'{\i}sica, Universidad de Sonora, 
% Apdo. Postal 1626, Hermosillo, Son.  83000, M\'exico.}
%\corresp[cor1]{barradas@fcfm.buap.mx}
%%%%%%%%%%%%%%%%%%%%%%%%%%%%%%%%%%%%%%%%%%%%%%%%%%%%%%%%%%%%%%%%%%%%%%%%%%%%%%%%%%%%%%%%%%%%%%%%%%%%%%%%%%%%%%%%%%%%
%%%%%%%%%%%%%%%%%%%%%%%%%%%%%%%%%%%%%%%%%%%%%%%%
%%%%%%%%%%%%%%%%%%%%%%%%%%%%%%%%%%%%%%%%%%%%%%

\allowdisplaybreaks \allowdisplaybreaks[2]
 \newcommand{\AddrFCFMBUAP}{
  Fac. de Cs. F\'{\i}sico Matem\'aticas, 
  Benem\'erita Universidad Aut\'onoma de Puebla,\\
  Apdo. Postal 1152, Puebla, Pue.  72000, M\'exico.}
%-----------------------------------------------------------------------------
 \newcommand{\AddrFCEBUAP}{
  Fac. de Cs. de la Electr\'onica, 
  Benem\'erita Universidad Aut\'onoma de Puebla,\\
  Apdo. Postal 542, Puebla, Pue. 72000, M\'exico.}
%-----------------------------------------------------------------------------
\newcommand{\AddrCINVESTAV}{
 Departamento de F\'{\i}sica, 
 Centro de Investigaci\'on y de Estudios Avanzados del Instituto Polit\'ecnico 
 Nacional, Apartado Postal 14-740, CDMX 07000, M\'exico.}
 %-----------------------------------------------------------------------------
\newcommand{\AddrUNISON}{
 Departamento de F\'{\i}sica, 
 Universidad de Sonora, Apdo. Postal 1626, Hermosillo, Son.  83000, M\'exico.}

\author{E. Barradas-Guevara}
 \email{barradas@fcfm.buap.mx}
 \affiliation{\AddrFCFMBUAP}
% 
%----------------------------------------------------------------------------- 
%
\author{O. F\'elix-Beltr\'an}
 \email{olga.felix@correo.buap.mx}
 \affiliation{\AddrFCEBUAP}
% 
%----------------------------------------------------------------------------- 
%
\author{F. Gonzalez-Canales}
 \email{felix.gonzalez@ific.uv.es}
 \affiliation{\AddrCINVESTAV}
% 
%----------------------------------------------------------------------------- 

%----------------------------------------------------------------------------- 
%
\author{E. Gonz\'alez-Hern\'andez}
 \email{}
 \affiliation{\AddrFCFMBUAP}
%
%----------------------------------------------------------------------------- 
%
\author{E. Rodr\'iguez-J\'auregui}
 \email{ezequiel.rodriguez@correo.fisica.uson.mx }
 \affiliation{\AddrUNISON}
% 
%----------------------------------------------------------------------------- 

%
\author{M. Zeleny-Mora}
 \email{barradas@fcfm.buap.mx}
 \affiliation{\AddrFCFMBUAP}

%%%%%%%%%%%%%%%%%%%%%%%%%%%%%%%%%%%%%%%%%%%%%%%%
%%%%%%%%%%%%%%%%%%%%%%%%%%%%%%%%%%%%%%%%%%%%%%%

\begin{abstract}
 In the theoretical framework of Two Higgs Doublet Model (2HDM) plus three 
 right-handed neutrinos we consider a universal treatment for the mass matrices, 
 aside from that the active neutrinos acquire their small mass through the type-I 
 seesaw mechanism. 
 Then, as long as a matrix with four-zero texture is used to represent the 
 right-handed neutrinos and Yukawa matrices, we obtain a unified treatment  where 
 all fermion mass matrices have four-zero texture.
 We obtain analytical and explicit expressions for the lepton flavour mixing 
 matrix PMNS in terms of fermion masses and parameters associated with the 
 2HDM-III.  
 Further, we compare these expressions of the PMNS matrix with the most up to date 
 values of masses and mixing in the lepton sector, via a likelihood test 
 $\chi^{2}$. We find that the analytical expressions that we derived reproduce 
 remarkably well the most recent experimental data of neutrino oscillations.  
\end{abstract}
\maketitle

%\vspace{1.5cm}

%\bigskip


%\tableofcontents

%%%%%%%%%%%%%%%%%%%%%%
\section{Introduction \hspace{13.6cm}}\label{sec:introduction}
%%%%%%%%%%%%%%%%%%%%
%
%
Although highly successful in terms of its phenomenological predictions, the Standard Model~(SM) of electroweak interactions seems incomplete from a theoretical view. In its present form, it is unable to predict the masses of fermions (leptons and quarks), or explain why there are several families of such particles. One of the most interesting phenomena is presented by the neutrino mixing, a phenomenon known as neutrino oscillation.
In concordance with the recent work focus on neutrino physics~\cite{Capozzi:2016rtj}, neutrino mass scale, corresponding Dirac or Majorana kind of fermion, and the source of Charge-Parity (CP) violation are unsolved questions. For that, see the experimental results concerning KamLAND (KL) reactor neutrinos~\cite{Gando:2010aa,Gando:2013nba,Seo:2014xei}, with respect to the expectations from reference Huber-M\"uller~(HM) spectra~\cite{Gando:2010aa,Gando:2013nba}. In each of the current high-statistics short-baseline(SBL) reactor experiments RENO~\cite{Seo:2014xei,RENO:2015ksa}, Double 
Chooz~\cite{Abe:2014bwa} and Daya Bay~\cite{An:2015nua}.
In general, if neutrinos are massive particles and their masses are non-degenerate, it is impossible to find a flavour basis in which the coincidence between flavour and mass eigenstates holds both for charged leptons and for neutrinos. Hence, the phenomenon of leptonic flavour mixing is naturally appear between three charged leptons and three massive neutrinos. If there exist irremovable phase factors in the 
Yukawa interactions, the CP violation will naturally appear both in the quark and lepton sector.

In this context, the flavour and mass generation are two concepts strongly 
intertwined. To know the flavour dynamic  in models beyond the SM, we need to understand the flavour mechanism  and mass generation  arising in the standard theory. In this theory, the Yukawa matrices are of great interest because the values of its elements define to the fermion masses, as well as its phases factors are related with the CP violation through the mixing matrix. 
%In this theory, Yukawa matrices are of great interest because its own values 
%defines the fermion masses. 

Moreover, the flavour changing currents arise from the not simultaneous diagonalization of  Higgs and Yukawa matrices. Particularly, we will study the flavour dynamics  through Yukawa matrices in the 2HDM-III (see therein references related with this model in Ref.~\cite{Felix-Beltran:2013tra}),  which into the processes comes with flavour violation through Higgs states, that is, it allows to appear the Flavour Changing Neutral Currents (FCNC) mediated by Higgs fields. 

Other models like the 2HDM-III allow the FCNC~\cite{Atwood:1996vj, Krawczyk:2007ne}. The difference between these models is in the
Yukawa structure and symmetries of the Higgs sector as well as the possible appearance of new sources of CP violation. In this work, the Higgs potential preserves the CP symmetry  with the Hermitian Yukawa matrices. 2HDM-III predicts three neutral states and a pair of charged states: $ H^0_{1,2,3}$ and $H^{\pm}_{1,2}$~\cite {Krawczyk:2005zy}.

In 2HDM-III, FCNC are kept under control by imposing some texture of Yukawa matrices that reproduce the observed fermion masses and mixing angles~\cite{Deppisch:2012vj}. Using texture forms allows for a direct relation between  the Yukawa matrix elements and the parameters related with the decay widths and cross section without losing 
the terms proportional to the light fermions masses. Specifically, considering a four-zero texture Yukawa matrix, one obtains in a natural way the Cheng-Sher ansatz for couplings flavour mix, which is widely used in the literature, where flavoured couplings are considered proportional to the involved fermion masses~\cite{Felix-Beltran:2013tra,Dorsner:2002wi}.

This work is realized in the frame of  2HDM-III, considering a hybrid treatment of the neutral leptonic sector through type-I seesaw mechanism. Moreover, a four-zero texture ansatz for Dirac and Majorana neutrino mass matrices, left and right-handed neutrinos respectively. We perform a statistical analysis of neutrino mixing angles using the likelihood test $\chi^{2}$.

%%%%%%%%%%%%%%%%%%%%%%
\section{~The~2HDM and seesaw mechanism \hspace{9.5cm} \label{sec:2HDM-III}}
%%%%%%%%%%%%%%%%%%%%%
%
%Yukawa Lagrangian in 2HDM-III allows changing flavour Higgs-fermions, this is,
In order to make a minimal extension of 2HDM by introducing right-handed neutrinos, we need to consider six neutrino fields; three left-handed neutrinos $\nu_{L}= \left( \nu_{eL}, \nu_{\mu L} , \nu_{\tau L} \right)^{\top}$ and three right-handed neutrinos $N_{R} = \left( N_{1R}, N_{2R}, N_{3R} \right)$. Where only the left-handed fields take part in the electroweak interactions.    
In context of Two Higgs Doublet Model plus massive neutrinos, 2HDM+3$\nu$, for Dirac leptons the Lagrangian of  Yukawa interactions has the form:
\begin{equation}
 {\cal L}_Y = 
 \sum_{k=1}^{2} 
 \left( \,
  \mathbf{Y}_{k}^{ \nu } \, \bar{L} \, \tilde{\Phi}_{k} \, N_{R} +  
  \mathbf{Y}_{k}^{l} \, \bar{L} \, \Phi_{k} \, l_{R} 
 \, \right) 
 + \textrm{H.c.} \, ,
\end{equation}
where $L = ( \nu_{l}, l^{-} )^{\top}_{L}$ is the left-handed doublet of $SU(2)$, the index $l$ represents the charged leptons. The $\Phi_{k} = ( \phi_{k}^{+}, \phi_{k}^{0} )^{\top}$ denotes the Higgs doublets with $\tilde{\Phi}_{k} = i \sigma_{2} \Phi_{k}^{*}$.  Finally, the $\mathbf{Y}_{k}^{j}$  with $j= l, \nu$, are the complex Yukawa 
$3\times3$ matrices.
In flavour space, the Dirac fermion mass matrix can be written as:
\begin{equation}\label{matriz de masa gral}
 \mathbf{M}_{j} = 
 \frac{1}{\sqrt{2}} 
 \left( \, 
    v_{1} \, \mathbf{Y}_{1}^{j} 
  + v_{2} \, \mathbf{Y}_{2}^{j} 
 \, \right),
\end{equation}
where $v_{1,2}$ are the vacuum expectation values (vev) associated with each of the Higgs doublets. In addition, these matrices can be diagonalized through a unitary transformation $\mathbf{U}$, such that:
\begin{equation}
 \mathbf{U}_{j\,L} \, \mathbf{M}_{j} \, \mathbf{U}_{j\,R}^{\dagger} = 
 \frac{1}{\sqrt{2}}
 \left( 
  v_{1} \, \mathbf{\tilde{Y}}_{1}^{j} + v_{2} \, \mathbf{\tilde{Y}}_{2}^{j}
 \right) = 
 %\Delta_{j}
 \diag \{\, m_{j1}, \, m_{j2}, \, m_{j3}\}
\end{equation}
where 
%$\Delta_{j} = \diag (\, m_{j1}, \, m_{j2}, \, m_{j3})$ and 
$\tilde{\bf Y}_{k}^{f} = {\bf U}_{jL} {\bf Y}_{k}^{j} {\bf U}_{jR}^{\dagger}$ are 
the Yukawa matrices in the mass basis, which give us the shape  of 
 Fermion-Fermion-Higgs couplings.

Here we consider that active neutrinos acquire their small mass through some seesaw 
mechanism. Hence, it is possible to write out the following hybrid mass term which involves both Dirac and Majorana neutrinos
\begin{equation}
\label{LMajoranaDirac}
 {\cal L}^{ D + M } = 
 - \overline{\nu}_{L} \, {\bf M}_{D} \, N_{R}
 - \frac{1}{2} \overline{\nu}_{L} \, {\bf M}_{L} \, \left( \nu_{L} \right)^{c}  
 - \frac{1}{2} \overline{ \left( N_{R} \right)^c} {\bf M}_{R} N_{R} 
 + \textrm{H.c.} \;.
\end{equation}
In the above expression $M_{D}$ is the Dirac neutrino mass matrix, 
while $M_{L}$ and $M_{R}$ are symmetric mass matrices because the corresponding 
mass terms are of the Majorana type.
%
%The mass terms we have seen before and a Majorana Lagrangian included in the theory 
%can give rise to the hybrid mass term associated to the neutrinos; since 
%theoretically nothing forbid it, {\it i.e.},  This term can be generated by the 
%hybrid Lagrangian ${\cal L}^{D+M}$ given as:
%This Lagrangian is the sum of the term left-handed Majorana mass, the Dirac mass 
%term and the left-handed Majorana mass, where 
%$\mathbf{M}^{M}_{L},\mathbf{M}^{M}_{R},\mathbf{M}^{D} $ are matrices of the order %$3 \times 3$; $\mathbf{M}^{M}_{L,R}$ are non-diagonal symmetric matrices and 
%$ \mathbf{M}^{D} $ is a non diagonal complex matrix,  
%$\nu^T_L= \left( \nu_{e L}, \nu_{\mu L}, \nu_{\tau L} \right) $
%
%and
%
%$\nu^{T}_R= \left( \nu_{e R},\\ \nu_{\mu R},\\ \nu_{\tau R} \right). $
%
%In the same way as for Majorana neutrinos, 
In this case the lepton number $L$ is not conserved.
In order to diagonalize the hybrid Lagrangian, Eq.~(\ref{LMajoranaDirac}), 
we can begin by rewriting to ${\cal L}^{ D + M }$ as follows:
%To diagonalize the Lagrangian~\ref{LMajoranaDirac} we can begin by rewriting it as 
%follows:
%
\begin{equation}
 {\cal L}^{D+M} = 
 - \frac{1}{2} \bar n_{L} {\bf M}^{ D+M } \left( n_{L} \right)^{c} 
 + \textrm{H.c.} \;,
\label{LMajoranaDiracsimple}
\end{equation}
where $ n_{L} = \left( \, \nu_{L} \, , \, \left( N_{R} \right)^{c} \, \right)$
and
\begin{equation}
 {\bf M}^{D+M} = 
 \left( \begin{array}{ccc}
  \mathbf{M}_L & 
  \mathbf{M}_{D} \\ \\
  \mathbf{M}_{D}^{\top} & 
  \mathbf{M}_R
 \end{array} \right)
 \label{matrizmdm}
\end{equation}
%
%$ \mathbf{M}^{D + M}$ 
is a $6\times 6$ complex symmetric matrix and can be presented in its diagonal form as:
\begin{equation}
 \mathbf{m} = 
 {\mathbf{U}}^{\top} 
 {\mathbf{M}}^{D+M}
 \mathbf{U}=
 \diag \{ \lambda_{1} ,\lambda_{2} \},
 \label{diaglambda12}
\end{equation}
where $\mathbf{U}$ is a $6\times 6$ unitary matrix.  
%y $ \mathbf{m}_{i k}= \mathbf{m}_{i} \delta_{i k}$ ($i=1,\cdots, 6$).
In the case that neutrino mass matrices satisfy the following hierarchy condition 
$ \mathbf{M}_{R}\gg \mathbf{M}_{D} \gg \mathbf{M}_{L}$, we obtain that eigenvalues 
of ${\mathbf{M}}^{D+M}$ matrix take the form:
\begin{equation}
 \lambda_{1} \approx {\bf M}_{R} 
 \quad \textrm{and} \quad
 \lambda_{2} \approx {\bf M}_{L} 
 - \mathbf{M}_{D} {\bf M}_{R}^{-1} \mathbf{M}_{D}^{\top}.
\end{equation}
The previous expression is known as type-(I+II) seesaw mechanism, and it is just the 
effective mass matrix of three active neutrinos.

%%%%%%%%%%%%%%%%%%%%%%%%%%%%%%%%
%\section{Neutrino masses and mixing matrices\label{sec:massmatrices} }
\section{~Fermion mass matrices \hspace{11.6cm} \label{sec:massmatrices} }
%%%%%%%%%%%%%%%%%%%%%%%%%%%%%%
%
%
In general, the Dirac fermion mass matrix has an arbitrary shape, while  the 
right-handed neutrino mass matrix must be symmetric, since these latter are Majorana 
particles.  
In particular, in this work we consider that, respectively, the Dirac fermion and 
right-handed neutrino mass matrices are represented with an Hermitian and complex 
symmetric matrix with a four-zero texture shape. The explicit form of these matrices 
are the following
%We consider the Dirac neutrinos mass matrix  $\mathbf{M}_{\nu{_{_D}}}$ likes 
%Hermitian  with a four-zero texture shape  as the   right-handed Majorana neutrinos 
%mass matrix $\mathbf{M}_{\nu{_{_R}}}$, which can be expressed as:
%
\begin{equation}\label{eq:MDmatrices}
 \mathbf{M}_{j} = 
 {\bf P}_{j}^{\dagger} \, \overline{\bf M}_{j} \, {\bf P}_{j} =
 \begin{pmatrix} \vspace{2mm}
  1 & 0 & 0 \\\vspace{2mm}
  0 & e^{-i \theta_C } & 0 \\ \vspace{2mm}
  0 & 0 & e^{-i \left( \theta_B + \theta_C \right) }  
 \end{pmatrix}
 \begin{pmatrix} \vspace{2mm} 
  0 &  \left | C_{j} \right| & 0 \\ \vspace{2mm}
  \left | C_{j} \right| & \widetilde{B}_{j} & \left | B_{j} \right| \\ \vspace{2mm}
  0 & \left | B_{j}\right|  & A_{j}
 \end{pmatrix} 
 \begin{pmatrix} \vspace{2mm}
  1 & 0 & 0 \\ \vspace{2mm}
  0 & e^{i \theta_C } & 0 \\ \vspace{2mm}
  0 & 0 & e^{i \left( \theta_B + \theta_C \right) }  
 \end{pmatrix}
 \qquad \textrm{and} \qquad 
 \mathbf{M}_{R} = 
 \begin{pmatrix}
  0 & c & 0\\
  c & \widetilde{b} & b\\
  0 & b & a
 \end{pmatrix},
\end{equation}
where $\theta_B \equiv  \arg \left \{ B_{j} \right \}$ and 
$\theta_C \equiv  \arg \left \{ C_{j} \right \}$.
%
%
%We are interested to assume the seesaw mechanism type I, this takes the hierarchy
%$ \mathbf{M}^{M}_{R}\gg \mathbf{M}^{D} \gg \mathbf{M}^{M}_{L}$. Also, in our case 
%we assume a hierarchical ansatz, which considers the Dirac neutrinos mass matrix 
%$\mathbf{M}^{D}$ and both Yukawa matrices $ \mathbf{Y}_{k}^{\nu_{l}}, \, (l=1,2)$, 
%as the same shape:
From the expressions for the Dirac fermion mass matrix given in 
Eqs.~(\ref{matriz de masa gral}) and~(\ref{eq:MDmatrices}), we obtain that 
$\mathbf{Y}_{k}^{j}$ Yukawa matrices also have a shape with four-zero texture, as 
shown below
\begin{equation}\label{conexion}
 \mathbf{M}_{j} = 
 \begin{pmatrix} \vspace{1mm}
  0 & C_{j} & 0 \\ \vspace{1mm}
  C_{j}^{\ast} & \widetilde{B}_{j} & B_{j} \\
  0 & B_{j}^{\ast} & A_{j}
 \end{pmatrix} = 
 \frac{ v \cos{\beta} }{ \sqrt{2} } 
 \left[
 \begin{pmatrix} \vspace{2mm}
  0 & C^{j}_{1} & 0\\ \vspace{2mm}
  C^{j \, \ast}_{1} & \widetilde{B}^{j}_{1} & B^{j}_{1} \\ \vspace{2mm}
  0 & B^{j \, \ast}_{1} & A^{j}_{1}
 \end{pmatrix} 
 + \tan{\beta}
 \begin{pmatrix} \vspace{2mm}
 0 & C^{j}_{2} & 0 \\ \vspace{2mm}
 C^{j \, \ast}_{2} & \widetilde{B}^{j}_{2} & B^{j}_{2} \\ \vspace{2mm}
 0 & B^{j \, \ast}_{2} & A^{j}_{2}
 \end{pmatrix} \right] \, ,
\end{equation}
where $\tan \beta = v_{2}/v_{1}$ and 
$v^2 = v_{1}^{2} + v_{2}^{2} = (246.22~\textrm{GeV})^{2}$.

Additionally, here we consider that the left-handed neutrinos acquire their small 
mass through the  type-I seesaw mechanism, which is defined as:
${\bf M}_{ \nu_{ L } } =  {\bf M}_D \; {\bf M}_{R}^{-1} \; {\bf M}_{D}^{\top}$.
%
%\begin{equation}\label{eq:matriz de masa efectiva} 
% {\bf M}_{ \nu_{ L } } =  
% {\bf M}^D \; \left( {\bf M}^{R} \right)^{-1} \; 
% \left( {\bf M}^{D} \right)^{\top}.
%\end{equation}
%
So, from the mass matrices given in Eq.~(\ref{eq:MDmatrices}) the $\mathbf{M}_{\nu_{L}}$ matrix takes the following explicit form
\begin{equation}
 \mathbf{M}_{ \nu_{L} } = 
 \mathbf{K} \, \overline{\mathbf{M}}_{ \nu_{L} } \, \mathbf{K} =
  \begin{pmatrix} \vspace{2mm}
  1 & 0 & 0 \\ \vspace{2mm}
  0 & e^{i \varphi_{B}/2 } & 0 \\ \vspace{2mm}
  0 & 0 & e^{i \varphi_{A}/2 }
 \end{pmatrix}
 \begin{pmatrix}\vspace{2mm}
  0 & \left| C_{ \nu_{L} } \right| & 0\\\vspace{2mm}
  \left| C_{ \nu_{L} } \right| & \left| \widetilde{B}_{ \nu_{L} } \right| & 
  \left| B_{ \nu_{L} } \right| \\ \vspace{2mm}
  0 &  \left| B_{ \nu_{L} } \right| & \left| A_{ \nu_{L} } \right|
 \end{pmatrix}
   \begin{pmatrix} \vspace{2mm}
  1 & 0 & 0 \\ \vspace{2mm}
  0 & e^{i \varphi_{B}/2 } & 0 \\ \vspace{2mm}
  0 & 0 & e^{i \varphi_{A}/2 }
 \end{pmatrix}, 
\end{equation}
where
\begin{equation}
 \begin{array}{l}\vspace{2mm}
 A_{ \nu_{L} } = \frac{ A_{D}^{2} }{ a }, \quad
 B_{ \nu_{L} } = \frac{ B_{D}^{\ast} C_{D}^{\ast} }{ c } 
  + A_{D} \left( \frac{ B_{D} }{ a } - \frac{ b C_{D}^{\ast} }{ a c } \right) 
  , \quad
 C_{ \nu_{L} } = \frac{ \left| C_{D} \right|^{2} }{ c }, \\
 \widetilde{B}_{ \nu_{L} } = 
  \left( - \frac{ b B_{D} }{ ac } 
   - \frac{ ( a \widetilde{b} - b^2 )C_{D}^{\ast}}{ ac^{2} } 
   + \frac{\widetilde{B}_{D} }{ c } \right) C_{D}^{\ast} 
   + \frac{\widetilde{B}_{D} C_{D}^{\ast}}{c} 
   + B_{D} \left( \frac{ B_{D} }{ a } - \frac{ b C_{D}^{\ast} }{ ac } \right).  
 \end{array}
\end{equation}
The elements of diagonal phase matrix $\mathbf{K}$ are defined as 
$\varphi_{A} \equiv \arg \left\{ A_{ \nu_{L} } \right\}$ and 
$\varphi_{B} \equiv \arg \left\{ \widetilde{B}_{ \nu_{L} } \right\}$. 
Also, the phase factors of $\mathbf{M}_{\nu_{L}}$ matrix must satisfy the conditions
$2 \arg \left\{ C_{ \nu_{L} } \right\} = \arg \left\{ \widetilde{B}_{ \nu_{L} } 
\right\}$ and
$2 \arg \left\{ B_{ \nu_{L} } \right\} = \arg \left\{ A_{ \nu_{L} } \right\} 
+ \arg \left\{ \widetilde{B}_{ \nu_{L} } \right\}$.

The real symmetric mass matrix $\overline{\bf M}_{f}$, with $f=u,d,l,\nu_{L}$, may 
be brought to diagonal form by means of an orthogonal transformation,
\begin{equation}
 \overline{\bf M}_{f} = 
 {\bf O}_{f} \; 
 \textrm{diag} \left( \lambda_{f1} \; , \lambda_{f2} \; , \lambda_{f3} \; \right) \; 
 {\bf O}_{f}^{\top}
\end{equation}
where the $\lambda_{f}$'s are the eigenvalues of ${\bf M}_{f}$ matrix and 
${\bf O}_{f}$ is a real orthogonal matrix. 
%Mainly, we express the entries of the matrices in terms of its invariants. 
Hence, the invariants of $\overline{\bf M}_{f}$ matrix are\footnote{In this 
expressions for the left-handed neutrinos $A_{f} = \left|A_{f} \right| $ and 
$\widetilde{B}_{f} = \left| \widetilde{B}_{f} \right|$.}
\begin{equation}
 \begin{array}{l}\vspace{2mm}
  \textrm{Tr} \left\{ \overline{\mathbf{M}}_{f} \right \} =
   A_{f} + \widetilde{B}_{f} = 
   \lambda_{f1} + \lambda_{f2} + \lambda_{f3}, \\ \vspace{2mm}
  \det\left\{ \overline{\mathbf{M}}_{f} \right \} =
   - A_{f} \left| C_{f} \right|^{2} = 
   \lambda_{f1}  \lambda_{f2}  \lambda_{f3}, \\ \vspace{2mm}
  \chi \left\{ \overline{\mathbf{M}}_{f} \right \} = \frac{1}{2} 
   \left( 
    \textrm{Tr} \left\{ \overline{\mathbf{M}}_{f}^{2} \right \} 
   - \textrm{Tr} \left\{ \overline{\mathbf{M}}_{f} \right \}^2 \right) =
   - A_{f} \widetilde{B}_{f} +|B_{f}|^{2}+|C_{f}|^{2} =
   - \lambda_{f1}  \lambda_{f2}  - \lambda_{f1} \lambda_{f3} 
   - \lambda_{f2}  \lambda_{f3}. 
 \end{array}
\end{equation}
%The corresponding $\mathbf{M}^R$ invariants have the same shape.
From the above expressions we may express the elements of $\overline{\bf M}_{f}$ 
matrices in terms of its mass eigenvalues. However, they are unable to give us 
information about the possible hierarchy in the mass spectrum. Therefore, a matrix 
with the four-zero texture shape  allows to have  a normal or inverted hierarchy in 
the fermionic masses. This latter hierarchy only is possible for the left-handed 
neutrino masses. 

%%%%%%%%%%%%%%%%%%%%%%%%%%%%%%%
\subsection{~{\it The mixing matrix as function of fermion masses} \hspace{7.3cm}} \label{subsec:PMNSmasses}
%%%%%%%%%%%%%%%%%%%%%%%%%%%%%
%
%
After obtaining the neutrino mass matrix through the type-I seesaw mechanism, let 
this matrix diagonalize in the context of two different scenarios, which depend on 
the mass hierarchy imposed on the neutrino mass matrix: 
Normal Hierarchy (NH) and Inverted Hierarchy (IH).

\begin{flushleft}
%%%%%%%%%%%%%%%%%%%
%\subsubsection{Normal Hierarchy \hspace{6cm}}
%%%%%%%%%%%%%%%%%%%
{\it Normal hierarchy}
\end{flushleft}
The NH in the eigenvalues of $ \mathbf{M}_{f}$ matrix is defined as 
$\lambda_{i3} > \lambda_{i2} > \lambda_{i1}$. Hence, the mass matrix parameters in 
terms of mass eigenvalues and the $(3,3)$ mass matrix entry, take the form 
%are expressed in function of these as
%
\begin{eqnarray} 
 \label{1jerarquia normal}
 \widetilde{B}_{f} & = & \lambda_{f1} + \lambda_{f2} + \lambda_{f3} - A_{f}, \\
 \label{2jerarquia normal}
 \left| C_{f} \right|^2 & = & 
  -\frac{\lambda_{f1} \lambda_{f2} \lambda_{f3} }{ A_{f} },\\
 \label{3jerarquia normal}
 \left| B_{f} \right|^2 & = & 
  \frac{(\lambda_{f3}-A_{f})(A_{f}-\lambda_{f1})(A_{f}-\lambda_{f2})}{A_{f}}.
\end{eqnarray}
According with the results, we have to take 
$\lambda_{fj}= - \left| \lambda_{fj} \right|$ with $j = 1,2,3$ such that 
\begin{equation}
 \begin{array}{lcl}\vspace{2mm}
  \lambda_{f3} > A_{i} > \lambda_{f2} & \textrm{for} & 
   \lambda_{f1}= - \left| \lambda_{f1} \right|, \\ \vspace{2mm}
  \lambda_{f3} > A_{i} > \lambda_{f1} & \textrm{for} &    
   \lambda_{f2}= - \left| \lambda_{f2} \right| ,\\ \vspace{2mm}
  \lambda_{f2} > A_{i} > \lambda_{f1} & \textrm{for} &
   \lambda_{f3}= - \left| \lambda_{f3} \right| .
 \end{array}
\end{equation}
In case of the charged leptons: 
$\lambda_{l1} = m_{e}, \; \lambda_{l2} = m_{\mu}, \; \lambda_{l3} = m_{\tau}$.
The NH is evident  by defining the adimensional parameters
$\widetilde{M}_{f} \equiv M_{f}/\lambda_{f3}$. Also, assuming this hierarchical 
ansatz, the heaviest particle is placed in the $(3,3)$ mass matrix entry. 
%whose expression is (the notation tilde corresponds to the NH)
%
%\begin{equation}
%\widetilde{M}_{i}=
%\begin{pmatrix}
% 0 & c_{i} & 0\\
% c_{i} & \widetilde{b}_{i} & b_{i}\\
% 0 & b_{i} & a_{i}
%\end{pmatrix}.
%\end{equation}
%
Then, it is  assumed that the parameter $a_{f} = A_{f} / \lambda_{f3} $ is very 
close to 1, therefore one can define $a_{f} \equiv 1 - \delta_{f}$, and the mass 
matrix takes the expression
\begin{eqnarray}
 \widetilde{M}_{f} & = & 
 \begin{pmatrix}
  0 & 
  \sqrt{ \frac{ \widetilde{\lambda}_{f1} \, \widetilde{\lambda}_{f2} }{ 
   1 - \delta_{f} } } & 0 \\
  \sqrt{ \frac{ \widetilde{\lambda}_{f1} \, \widetilde{\lambda}_{f2} }{ 
   1 - \delta_{f} } } & 
  \widetilde{\lambda}_{f1} - \widetilde{\lambda}_{f2} + \delta_{i} & 
  \sqrt{ \frac{ \delta_{f} }{ 1 - \delta_{f} } \xi_{f1} \xi_{f2}} \\
  0 & \sqrt{ \frac{ \delta_{f} }{ 1 - \delta_{f} } \xi_{f1} \xi_{f2}} & 
  1 - \delta_{f}
 \end{pmatrix},
\end{eqnarray}
where
\begin{equation}
 \xi_{f1} = \left(1 - \delta_{i} - \widetilde{\lambda}_{f1} \right) 
 \quad \textrm{and} \quad
 \xi_{f2} = \left(1 - \delta_{i} + \widetilde{\lambda}_{f2} \right),
\end{equation}
with $\widetilde{\lambda}_{f1} = \lambda_{f1}/\lambda_{f3}$ and 
$\widetilde{\lambda}_{f2} = \left| \lambda_{f2} \right| / \lambda_{f3}$.
%
%The orthogonal rotation matriz $\widetilde{O}_i$ is taken as
%
%\begin{equation}\label{matriz ortogonal real normal}
% \widetilde{O}_i = 
%\left(|\widetilde{M}_{i1}\rangle, |\widetilde{M}_{i2}\rangle, 
%\widetilde{M}_{i3}\rangle\right).
%\end{equation}
%

%%%%%%%%%%%%%%%%%%%
%\subsubsection{Inverted Hierarchy}
%%%%%%%%%%%%%%%%%%%
\begin{flushleft}
 {\it Inverted hierarchy}
\end{flushleft}
For an inverted hierarchy (IH), the relation between the eigenvalues is 
$\lambda_{f2} > \lambda_{f1} > \lambda_{f3}$. Analogous to NH, the mass matrix 
parameters are expressed in terms of eigenvalues as
\begin{eqnarray} 
 \label{1jerarquia invertida}
  \widetilde{B}_{f} & = & \lambda_{f1} + \lambda_{f2} + \lambda_{f3} - A_{f},  \\
 \label{2jerarquia invertida}
  \left| C_{f} \right|^2 & = & 
   -\frac{ \lambda_{f1} \lambda_{f2} \lambda_{f3} }{ A_{f} }, \\
 \label{3jerarquia invertida}
  \left| B_{f} \right|^2 & = & 
   \frac{( A_{f} - \lambda_{f3} ) ( A_{f} - \lambda_{f1} ) 
   (\lambda_{f2} - A_{f}) }{ A_{f} }.
\end{eqnarray}
According with the results, we have to take 
$\lambda_{f j}= -|\lambda_{f j}|$ with $j=1,2,3$ such that
\begin{equation}
 \begin{array}{lcl}\vspace{2mm}
  \lambda_{f2} > A_{i} > \lambda_{f3} & \textrm{for} & 
   \lambda_{f1}= - \left| \lambda_{f1} \right|, \\ \vspace{2mm}
  \lambda_{f1} > A_{i} > \lambda_{f3} & \textrm{for} &    
   \lambda_{f2}= - \left| \lambda_{f2} \right|, \\ \vspace{2mm}
  \lambda_{f2} > A_{i} > \lambda_{f1} & \textrm{for} &
   \lambda_{f3}= - \left| \lambda_{f3} \right| .
 \end{array}
\end{equation}
%$\lambda_{i2}>A_{i}>\lambda_{i3},\, \lambda_{i1}>A_{i}>\lambda_{i3},\, %\lambda_{i2}>A_{i}>\lambda_{i1}$ respectively.
%
For neutrinos: 
$\lambda_{ \nu_{L 1} } = m_{ \nu{_1} }$, 
$\lambda_{ \nu_{L 2} } = m_{ \nu{_2} }$, 
$\lambda_{ \nu_{L 3} } = m_{ \nu{_3} }$; 
and for the charged leptons: 
$\lambda_{l1} = m_{e}$, $\lambda_{l2} = m_{\mu}$, $\lambda_{l3} = m_{\tau}$. 
For this hierarchy, the mass matrix is 
%(the notation hat corresponds to the IH):
%
\begin{eqnarray}
 \widetilde{M}_{i} = 
 \begin{pmatrix}
  0 & \sqrt{ \frac{ \widetilde{\lambda}_{f1} \, \widetilde{\lambda}_{f3} }{ 
   1 - \delta_{f} } } & 0 \\
  \sqrt{ \frac{\widetilde{\lambda}_{f1} \, \widetilde{\lambda}_{f3} }{ 
   1 - \delta_{f} } } & 
  - \widetilde{\lambda}_{f1} + \widetilde{\lambda}_{f3} + \delta_{f} & 
  \sqrt{ \frac{ \delta_{f} }{ 1 - \delta_{f} } \xi_{f1} \xi_{f3} } \\
  0 & \sqrt{ \frac{ \delta_{f} }{ 1 - \delta_{f} } \xi_{f1} \xi_{f3} }  & 
  1 - \delta_{f}
 \end{pmatrix}
\end{eqnarray}
where
\begin{equation}
 \xi_{f1} = \left( 1 - \delta_{f} + \widetilde{\lambda}_{f1} \right), 
 \quad \textrm{and} \quad
 \xi_{f3} = \left( 1 - \delta_{f} - \widetilde{\lambda}_{f3} \right),
\end{equation}
with $\widetilde{\lambda}_{f3} = \lambda_{f3}/\lambda_{f2}$ and 
$\widetilde{\lambda}_{f1} = \left| \lambda_{f1} \right| / \lambda_{f2}$.
%The orthogonal rotation matriz $\widehat{O}_i$ is as
%
%\begin{equation} \label{matriz ortogonal real invertida}
%   \widehat{O}_i=\left(-|\widehat{M}_{i1}\rangle, |\widehat{M}_{i2}\rangle, |
%\widehat{M}_{i3}\rangle\right).
%\end{equation}
%

For a normal~[inverted] hierarchy in the neutrino mass spectrum the real orthogonal 
matrix that diagonalized the fermion mass matrix with four-zero texture, in terms of 
fermion masses has the form:
\begin{equation}\label{eq:fmatrix}
 {\bf O}_{f} = 
 \left( \begin{array}{ccc} \vspace{2mm}
  \sqrt{ \frac{ \widetilde{m}_{f 2[1] } \, \xi_{f 1[3] } }{ {\cal D}_{f 1[3] } } } & 
  -\sqrt{ \frac{ \widetilde{m}_{f 1[3] } \, \xi_{f 2[1] } }{ {\cal D}_{f 2[1]} } } &
  \sqrt{ \frac{ \widetilde{m}_{f 1[3]} \, \widetilde{m}_{f 2[1]} \, \delta_{f} }{ 
   {\cal D}_{f 3[2]} } } \\ \vspace{2mm}
  \sqrt{ \frac{ \widetilde{m}_{f 1 [3]} \, \left( 1 - \delta_{f} \right) \, 
   \xi_{f 1[3]} }{ {\cal D}_{f 1[3]} } } &
  \sqrt{ \frac{ \widetilde{m}_{f 2 [1]} \, \left( 1 - \delta_{f} \right) \, 
   \xi_{f 2[1]} }{ {\cal D}_{f 2[1]} } } &
  \sqrt{ \frac{ \delta_{f} \, \left( 1 - \delta_{f} \right) }{ {\cal D}_{f 3[2]} } } 
   \\ \vspace{2mm}
  - \sqrt{ \frac{ \widetilde{m}_{f 1[3]} \, \delta_{f} \, \xi_{f 2[1]} }{ 
   {\cal D}_{f 1[3]} } }  &
  - \sqrt{ \frac{ \widetilde{m}_{f 2[1]} \, \delta_{f} \, \xi_{f 1[3]} }{ 
   {\cal D}_{f 2[1]} } }  &
  \sqrt{ \frac{ \xi_{f 1[3]} \, \xi_{f 2[1]} }{ {\cal D}_{f 3[2]} } } 
 \end{array}  \right) .
\end{equation}
In this matrix we have
\begin{equation}
 \begin{array}{l} \vspace{2mm}
  \xi_{f 1[3]} = 
   1 - \widetilde{m}_{f 1 [3]} - \delta_{f},\quad
  \xi_{f 2[1]} = 
   1 + \widetilde{m}_{f 2 [1]} - \delta_{f},\\ \vspace{2mm}
  {\cal D}_{f 1[3]} =  
   \left( 1 - \delta_{f} \right) 
   \left( \widetilde{m}_{f 1 [3]} + \widetilde{m}_{f 2 [1]} \right) 
   \left( 1 - \widetilde{m}_{f 1 [3]} \right),\\ \vspace{2mm}
  {\cal D}_{f 2[1]} = 
   \left( 1 - \delta_{f} \right) 
   \left( \widetilde{m}_{f 1 [3]} + \widetilde{m}_{f 2 [1]} \right) 
   \left( 1 + \widetilde{m}_{f 2 [1]} \right),\\ \vspace{2mm}
  {\cal D}_{f 3[2]} = 
   \left( 1 - \delta_{f} \right) \left( 1 - \widehat{m}_{f 1 [3]} \right) 
   \left( 1 + \widehat{m}_{f 2 [1]} \right) .
 \end{array}
\end{equation}
Now the subindex $f$ is considering as $f = u,d, \nu, l$. 
From Eqs.~(\ref{conexion}) and~(\ref{eq:fmatrix}) we obtain that the elements of the 
Yukawa matrices in the base of the mass $\widetilde{\bf Y}_{k}^{f}$ obey the called 
Cheng and Sher relation~\cite{Felix-Beltran:2013tra}
\begin{equation}
 \left( \widetilde{\bf Y}_{k}^{j} \right)_{ \texttt{kl} } = 
 \frac{ \sqrt{ m_{j\texttt{k}} \; m_{j\texttt{l}} } }{ v }
 \left( \widetilde{\bf \chi}_{k}^{j} \right)_{ \texttt{kl} } ,
\end{equation} 
where $\texttt{k,l} = 1, 2, 3$ and 
$\left( \widetilde{\bf \chi}_{k}^{j} \right)_{ \texttt{kl} } $ are complex functions
of the Yukawa matrix parameters and the mass matrix parameter $\delta_{j}$ which is associated with the 2HDM.
\begin{flushleft}
 {\it The flavour mixing matrix}
\end{flushleft}
The flavour mixing matrix of leptons, $\vpmns$  arises from the lack of 
correspondence between the diagonalization of the mass matrices of the charged 
leptons and left-handed neutrinos, and this is defined as:
\begin{equation}
 \vpmns= {\mathbf{U}_{l}}^\dag \mathbf{U}_{\nu} 
 \qquad \textrm{with} \qquad 
 {\mathbf{U}}_{\nu,l}={\mathbf{P}}_{\nu,l}  {\mathbf{ O}}_{\nu,l}.
\end{equation}
Also, the lepton mixing matrix can be written as:
\begin{equation}\label{V}
 \vpmns = {\bf O}_{l}^{\top} {\bf P}^{ \nu - l } {\bf O}_{\nu},
\end{equation}
where 
${\bf P}^{ \nu - l } = \textrm{diag} 
\left( 1 , e^{ i \Phi_{1} }, e^{ i \Phi_{2} }\right)$ with the phases factors
$\Phi_{1} = \varphi_{B}/2 - \theta_{C} $ and
$\Phi_{2} = \varphi_{A}/2 - \theta_{B} - \theta_{C}$. 
Finally, the theoretical entries of the matrix $ \vpmns $ for the NH~[IH] are given 
as:
%m_{\nu_{1[1]}
\begin{eqnarray}\label{elementos de matriz}
 V^{th}_{e 1} = &
  \sqrt{ 
   \frac{ \widetilde{m}_{\mu} \, \widetilde{m}_{\nu 2 [1]} \, 
    \xi_{l 1} \, \xi_{\nu 1 [3]} }{ D_{l1} \, D_{\nu 1[3]}} } 
  + \sqrt{ \frac{ \widetilde{m}_{e} \, \widetilde{m}_{\nu 1[3] } }{ 
    D_{l1} \, D_{\nu 1[3]}} }  
  \left( \sqrt{ \left( 1 - \delta_{\nu} \right) \left( 1 - \delta_{l} \right) 
   \xi_{l1} \, \xi_{\nu1 [3] } } \, e^{i \Phi _1} 
   + \sqrt{ \delta _{\nu } \, \delta_{l} \, \xi_{l2} \, \xi_{\nu 2[1] } } \, 
   e^{i \Phi _2} \right), \nonumber\\
 V^{th}_{e 2} = &
  - \sqrt{ \frac{ \widetilde{m}_{\mu } \, \widetilde{m}_{\nu 1 [3]} \, 
   \xi_{l1} \, \xi_{\nu 2 [1]}   }{ D_{l1} \, D_{ \nu 2[1] } } }  
  + \sqrt{ \frac{ \widetilde{m}_{e} \, \widetilde{m}_{\nu 2 [1]} }{ 
   D_{l1} \, D_{\nu 2 [1]} } } 
  \left( \sqrt{ \left( 1 - \delta _{\nu} \right) \left( 1 - \delta_{l} \right) 
   \xi_{l1} \, \xi_{\nu 2 [1]} }  e^{i \Phi_1} 
   + \sqrt{ \delta_{\nu} \, \delta_{l} \, \xi_{l2} \, \xi_{\nu 1 [3]} } 
    e^{i \Phi_2} \right), \nonumber \\
 V^{th}_{e 3} = &
  \sqrt{ \frac{ \widetilde{m}_{\mu } \, \widetilde{m}_{\nu 1[3]} \, 
   \widetilde{m}_{\nu 2[1]} \, \delta_{\nu} \, \xi_{l1} }{ 
   D_{l1} \, D_{\nu 3[2]}} } 
  + \sqrt{ \frac{\widetilde{m}_{e} }{ D_{l1} \, D_{\nu 3[2]} } } 
  \left( \sqrt{ \delta_{\nu} \left( 1 - \delta_{\nu } \right)  
    \left( 1 - \delta_{l} \right) \xi_{l1} } e^{i \Phi_1} 
   - \sqrt{ \delta_{l} \, \xi_{l2} \, \xi_{\nu 1[3]} \, \xi_{\nu 2[1]} } 
    e^{i \Phi_2} \right) , \nonumber \\
 V^{th}_{\mu 1} = &   
  - \sqrt{ \frac{ \widetilde{m}_{e} \, \widetilde{m}_{\nu 2 [1]} \, 
    \xi_{l 2} \, \xi_{\nu 1 [3]} }{ D_{l2} \, D_{\nu 1[3]}} } 
  + \sqrt{ \frac{ \widetilde{m}_{\mu} \, \widetilde{m}_{\nu 1[3] } }{ 
    D_{l2} \, D_{\nu 1[3]}} }  
  \left( \sqrt{ \left( 1 - \delta_{\nu} \right) \left( 1 - \delta_{l} \right) 
   \xi_{l2} \, \xi_{\nu1 [3] } } \, e^{i \Phi _1} 
   + \sqrt{ \delta _{\nu } \, \delta_{l} \, \xi_{l1} \, \xi_{\nu 2[1] } } \, 
   e^{i \Phi _2} \right), \nonumber\\
  V^{th}_{\mu 2} = &
   \sqrt{ \frac{ \widetilde{m}_{e} \, \widetilde{m}_{\nu 1 [3]} \, 
   \xi_{l2} \, \xi_{\nu 2 [1]}   }{ D_{l2} \, D_{ \nu 2[1] } } }  
   + \sqrt{ \frac{ \widetilde{m}_{\mu} \, \widetilde{m}_{\nu 2 [1]} }{ 
    D_{l2} \, D_{\nu 2 [1]} } } 
   \left( \sqrt{ \left( 1 - \delta _{\nu} \right) \left( 1 - \delta_{l} \right) 
    \xi_{l2} \, \xi_{\nu 2 [1]} }  e^{i \Phi_1} 
    + \sqrt{ \delta_{\nu} \, \delta_{l} \, \xi_{l1} \, \xi_{\nu 1 [3]} } 
     e^{i \Phi_2} \right), \\  
 V^{th}_{\mu 3} = &
  - \sqrt{ \frac{ \widetilde{m}_{e} \, \widetilde{m}_{\nu 1[3]} \, 
   \widetilde{m}_{\nu 2[1]} \, \delta_{\nu} \, \xi_{l2} }{ 
   D_{l2} \, D_{\nu 3[2]}} } 
  + \sqrt{ \frac{\widetilde{m}_{\mu} }{ D_{l2} \, D_{\nu 3[2]} } } 
  \left( \sqrt{ \delta_{\nu} \left( 1 - \delta_{\nu } \right)  
    \left( 1 - \delta_{l} \right) \xi_{l2} } e^{i \Phi_1} 
   - \sqrt{ \delta_{l} \, \xi_{l1} \, \xi_{\nu 1[3]} \, \xi_{\nu 2[1]} } 
    e^{i \Phi_2} \right) , \nonumber \\  
 V^{th}_{\tau 1} = &
  \sqrt{ \frac{ \widetilde{m}_{e} \, \widetilde{m}_{\mu} \, 
   \widetilde{m}_{\nu 2[1]} \, \delta_{l} \, \xi_{\nu 1[3]} }{ 
   D_{l3} \, D_{\nu 1[3]}} } 
  + \sqrt{ \frac{\widetilde{m}_{\nu 1[3]} }{ D_{l3} \, D_{\nu 1[3]} } } 
  \left( \sqrt{ \delta_{l} \left( 1 - \delta_{\nu } \right)  
    \left( 1 - \delta_{l} \right) \xi_{\nu 1[3]} } e^{i \Phi_1} 
   - \sqrt{ \delta_{\nu} \, \xi_{l1} \, \xi_{l2} \, \xi_{\nu 2[1]} } 
    e^{i \Phi_2} \right) , \nonumber \\  
 V^{th}_{\tau 2} = &
  - \sqrt{ \frac{ \widetilde{m}_{e} \, \widetilde{m}_{\mu} \, 
   \widetilde{m}_{\nu 1[3]} \, \delta_{l} \, \xi_{\nu 2[1]} }{ 
   D_{l3} \, D_{\nu 2[1]}} } 
  + \sqrt{ \frac{\widetilde{m}_{\nu 2[1]} }{ D_{l3} \, D_{\nu 2[1]} } } 
  \left( \sqrt{ \delta_{l} \left( 1 - \delta_{\nu } \right)  
    \left( 1 - \delta_{l} \right) \xi_{\nu 2[1]} } e^{i \Phi_1} 
   - \sqrt{ \delta_{\nu} \, \xi_{l1} \, \xi_{l2} \, \xi_{\nu 1[3]} } 
    e^{i \Phi_2} \right) , \nonumber \\     
 V^{th}_{\tau 3} = &
  \sqrt{ \frac{ \widetilde{m}_{e} \, \widetilde{m}_{\mu} \, 
   \widetilde{m}_{\nu 1[3]} \, \widetilde{m}_{\nu 2[1]} \, 
   \delta_{l} \, \delta_{\nu} }{ D_{l3} \, D_{\nu 3[2]}} } 
  + \frac{1 }{ \sqrt{ D_{l3} \, D_{\nu 3[2]} } } 
  \left( \sqrt{ \delta_{l} \,  \delta_{\nu } 
   \left( 1 - \delta_{\nu } \right) \left( 1 - \delta_{l} \right) } e^{i \Phi_1} 
   - \sqrt{ \xi_{l1} \, \xi_{l2} \, \xi_{\nu 1[3]} \xi_{\nu 2[1]} } 
    e^{i \Phi_2} \right) .   \nonumber         
\end{eqnarray}
\subsection{~{\it The symmetric parameterization} \hspace{10.1cm} \label{subsec:anglephase}}
%%%%%%%%%%%%%%%%%%%%%%%%%%%%%%%%%%
%
%
In the basis where flavour eigenstates of three charged leptons are identified with their mass eigenstates, the flavour eigenstates of three neutrinos can be written as 
\begin{equation}\label{Eq:pmns:comp}
 %\vpmns =
 \begin{pmatrix}
  \nu_{e} \\
  \nu_{\mu} \\
  \nu_{\tau}
 \end{pmatrix} =
 \begin{pmatrix}
  V_{e 1}    & V_{e 2}    & V_{e 3} \\
  V_{\mu 1}  & V_{\mu 2}  & V_{\mu 3} \\
  V_{\tau 1} & V_{\tau 2} & V_{\tau 3} 
 \end{pmatrix}
  \begin{pmatrix}
  \nu_{1} \\
  \nu_{2} \\
  \nu_{3}
 \end{pmatrix}.
\end{equation}
As neutrinos are Majorana particles, the nine elements of PMNS lepton mixing matrix 
can be parameterized by using three rotation angles and three CP-violating 
phases~\cite{Fritzsch:1999ee}. 
%considered in the angle-phase parameterization of the lepton mixing matrix, this 
%one can be represented in terms of three angles and three 
%phases~\cite{Fritzsch:1999ee}. 
In the so called symmetrical parametrization, the mixing matrix has the 
shape~\cite{Schechter:1980gr,Rodejohann:2011vc}:
\begin{equation}\label{eq:symmetric_para}
 \vpmns = 
 \left( 
 \begin{array}{ccc}
  \text{c}_{12} \text{c}_{13} &  
  \text{s}_{12} \text{c}_{13} e^{ - i \phi_{12} } & 
   \text{s}_{13} e^{ - i \phi_{13} } \\
  - \text{s}_{12} \text{c}_{23} e^{ i \phi_{12} } 
   - \text{c}_{12} \text{s}_{13} \text{s}_{23} e^{-i( \phi_{23} - \phi_{13} ) } & 
  \text{c}_{12} \text{c}_{23} 
   - \text{s}_{12} \text{s}_{13} \text{s}_{23} 
   e^{ - i ( \phi_{23} + \phi_{12} - \phi_{13} ) } & 
  \text{c}_{13} \text{s}_{23} e^{- i \phi_{23} } \\
  \text{s}_{12} \text{s}_{23} e^{ i ( \phi_{23} + \phi_{12} ) } 
   - \text{c}_{12} \text{s}_{13} \text{c}_{23} e^{ i \phi_{13} } &
  - \text{c}_{12} \text{s}_{23} e^{ i \phi_{23} } 
   - \text{s}_{12} \text{s}_{13} \text{c}_{23} 
   e^{ -i ( \phi_{12} - \phi_{13} ) } & \text{c}_{13} \text{c}_{23} 
 \end{array} \right),
\end{equation}
where $\textrm{c}_{ij} = \cos \theta_{ij}$ and $\textrm{s}_{ij} = \sin \theta_{ij}$. 
In this parametrization, the relation between flavour mixing angles and the entries of ${\bf V}_{\rm PMNS}$ matrix is
\begin{equation}\label{Eq:senoscuadrados}
 \sin^{2} \theta_{13} \equiv \left| V_{e 3} \right|^{2}, \quad 
 \sin^{2} \theta_{12} \equiv \frac{\left| V_{e 2}   \right|^{2} }{ 
  1 - \left| V_{e 3} \right|^{2}}, \quad
 \sin^{2} \theta_{23} \equiv \frac{\left| V_{\mu 3} \right|^{2} }{ 
  1 - \left| V_{e 3} \right|^{2}}.
\end{equation}
From the above expressions for the mixing angles, we can conclude that these are 
exactly the same expressions that are obtained in the Standard 
parametrization~\cite{Agashe:2014kda}. 
%
%From the comparing the above expressions with those obtained in the standard 
%parametrization~\cite{Agashe:2014kda}, we conclude that are exactly the same 
%expressions. 
In fact, the difference between the symmetric and standard parametrization is 
explicitly manifest in the CP invariants. The Jarlskog invariant which is used for describing the CP violation in conventional neutrino oscillations is defined as: $J_{\rm CP} = {\cal I}m \left \{ V_{e1}^{*} V_{\mu 3}^{*} V_{e3} V_{\mu 1} \right \} $. 
%

%From the phase invariants given in Eqs.~\eqref{INV:JCP} and~\eqref{INV:I1-I2} the equivalence between the PDG and the symmetrical parameterization may be expressed as 
%${\bf U}_{ \text{PDG} } = {\bf K} {\bf U}_{ \text{Sym} }$, where 
%${\bf K} = \textrm{diag} \left \{ 1, e^{i \frac{ \alpha_{21} }{2} }, e^{i \frac{ \alpha_{31} }{2} } \right \}$
%with $\delta_{CP} = \phi_{13} - \phi_{23} - \phi_{12}$, $\alpha_{21} = - 2 \phi_{12}$ and 
%$\alpha_{31} = - 2 ( \phi_{12} + \phi_{23} )$.
%

%%%%%%%%%%%%%%%%%%%%%%%%%%%%%
\section{ Numerical analysis \hspace{12.3cm} \label{sec:numericalanalysis}}
%%%%%%%%%%%%%%%%%%%%%%%%%%%%
%
In this section we make a likelihood test $\chi^{2}$ with the purpose of obtaining 
the best fit point~(BFP), which allows us to get  the numerical values of some free 
parameters in the $\chi^{2}$ function. But before, we can take advantage of the last 
exprimental data reported by Planck collaboration~\cite{Ade:2015xua} and global 
fits of neutrino oscillations data~\cite{Forero:2014bxa}. All this in order to 
reduce the degrees of freedom in the analysis. 

%In this work, we minimize the function $\chi^{2}$  and obtain the best fit point 
%(BFP). From that, we fixed three free parameters of $\chi^{2}$. So, our 
%likelihood analysis has just one degree of freedom. 

%
%%%%%%%%%%%%%%%%%%%%%%%%%%%%%%%%
\subsection{ {\it Neutrino mass bounds} \hspace{11.8cm} \label{subsec:neutrinosmass}}
%%%%%%%%%%%%%%%%%%%%%%%%%%%%%%%%
%
In the three flavour context there are six independent parameters which govern the 
behaviour of neutrino oscillations: the differences of the squared neutrino masses,
flavour mixing angles and the Dirac CP-violating phase. The definition of first one is 
%One of the main parameters that characterize to the nuetrino oscillations is the 
%differences of the squared neutrino masses, whose definition is  
$\Delta m_{i j}^{2} \equiv m^{2}_{ \nu_{i} } - m^{2}_{ \nu_{j} }$.  
For an normal~[inverted]~hierarchy in the neutrino mass spectrum, we can express 
two of the neutrino masses  in terms of the heaviest neutrino mass, as 
well as $\Delta m_{i j}^{2}$ parameter, as:
\begin{equation}\label{Eq:IH}
 \begin{array}{l}\vspace{2mm}
  m_{ \nu_{1 [3]} } = 
   \sqrt{ m^{2}_{\nu_{3 [2]} } - \Delta m^{2}_{31 [23]} } 
  \quad~\textrm{and}~\quad
  m_{ \nu_{2 [1]} } = 
   \sqrt{ m^{2}_{\nu_{3 [2]} } - \Delta m^{2}_{32 [21]} }.
 \end{array}
\end{equation}
The heavy neutrino mass must satisfy the relation 
$m_{ \nu_{3} }^{2} \geq \Delta m^{2}_{31 [23] }$,
and can be considered like the only one free parameter in the above relations, 
since the oscillation parameters~$\Delta m_{i j}^{2}$ are experimentally determined.
The values for the parameters $\Delta m_{i j}^{2}$ at BFP$\pm 1 \sigma$, 
$2 \sigma$ and $3\sigma$ reported in Ref.~\cite{Forero:2014bxa} are:
\begin{equation}\label{Eq:Val_DMij}
 \begin{array}{rl}\vspace{2mm}
  \Delta m_{21}^{2} \; \left( 10^{-5} \textrm{eV}^{2} \right)  = & 
   7.60_{-0.18}^{+0.19}, \; 7.26-7.99, \; 7.11-8.18, \\ \vspace{2mm}
  \left| \Delta m_{31}^{2} \right| \; \left( 10^{-3} \textrm{eV}^{2} \right) = & 
  \left \{ 
   \begin{array}{l}\vspace{2mm}
    2.48_{-0.07}^{+0.05}, \; 2.35-2.59, \; 2.30-2.65, \\ 
    2.38_{-0.06}^{+0.05}, \; 2.26-2.48, \; 2.20-2.54.
   \end{array}    \right.
 \end{array}
\end{equation}
In the above expressions for the parameter $\Delta m_{31}^{2}$  
the upper~[lower] row correspond to the values for a normal~[inverted] hierarchy 
in the mass spectrum. Moreover, the sum of the mass of the active neutrinos must 
comply with inequality; $\sum m_{\nu_{i}} < 0.23$, for the following actual 
number of active neutrinos $N_{\textrm eff} = 3.15 \pm 0.23$~\cite{Ade:2015xua}. 
These results are independent of the hierarchy of the neutrino mass spectrum. 
From Eqs.~(\ref{Eq:IH}) and~(\ref{Eq:Val_DMij}) the allowed ranges for the neutrino 
masses are obtained and given in the Table~\ref{Tab:masas_nu}. 
%are given by the ranges of theoretical masses values. 
Also it is easy to conclude that for both hierarchies, there is the possibility that 
the lightest neutrino could be a massless particle.

%Moreover, one computes these mass ranges for 
%$\textrm{BFP}\pm 1\sigma,\, \textrm{BFP}\pm2\sigma$ and 
%$\textrm{BFP}\pm 3\sigma$. Results are shown in Table~\ref{Tab:masas_nu}.
%
\begin{table}[!htbp]
 \centering
 \begin{tabular}{lcccc} \hline 
  {\footnotesize {\bf Hierarchy}} & 
   {\footnotesize $m_{\nu_1}~(10^{-2}$eV)} & 
   {\footnotesize $m_{\nu_2}~(10^{-2}$eV)} & 
   {\footnotesize $m_{\nu_3}~(10^{-2}$eV)} &   
   {\footnotesize $\Delta m_{ij}^{2}$ (eV)} \\ \hline
  \multirow{4}{*}{{\bf Normal}} & 
   {\footnotesize $[0 , 7.12]$ } & 
   {\footnotesize $[8.72 \times 10^{-1} , 7.18]$ } & 
   {\footnotesize $[4.98 , 8.69]$ } & 
   {\footnotesize BFP } \\  & 
   {\footnotesize $[0 , 7.18]$ } & 
   {\footnotesize $[8.61 \times 10^{-1},  7.23]$ } & 
   {\footnotesize $[4.91 , 8.71]$ } & 
   {\footnotesize BFP$\pm 1\sigma$ } \\  & 
   {\footnotesize $[0 , 7.25]$ } & 
   {\footnotesize $[8.51 \times 10^{-1} , 7.30]$ } & 
   {\footnotesize $[4.84 , 8.74]$ } & 
   {\footnotesize BFP$\pm2\sigma$ } \\  & 
   {\footnotesize $[0 , 7.32]$ } & 
   {\footnotesize $[8.40 \times 10^{-1} , 7.37]$ } & 
   {\footnotesize $[4.76 , 8.76]$ } & 
   {\footnotesize BFP$\pm3\sigma $ } \\ \hline
  \multirow{4}{*}{{\bf Inverted}} & 
   {\footnotesize $[4.87 , 8.19]$ } & 
   {\footnotesize $[4.96 , 8.23]$ } & 
   {\footnotesize $[0 , 6.58]$ } & 
   {\footnotesize BFP } \\  & 
   {\footnotesize $[4.81 , 8.21]$} & 
   {\footnotesize $[4.89 , 8.24]$} & 
   {\footnotesize $[0 , 6.64 ]$} & 
   {\footnotesize BFP$\pm 1\sigma $} \\  & 
   {\footnotesize $[4.75 , 8.22]$} &
   {\footnotesize $[4.83 , 8.26]$} & 
   {\footnotesize $[0 , 6.70]$} & 
   {\footnotesize BFP$\pm2\sigma $} \\  & 
   {\footnotesize $[4.69 , 8.23]$} & 
   {\footnotesize $[4.76 , 8.27]$} & 
   {\footnotesize $[0 , 6.76]$} & 
   {\footnotesize BFP$\pm3\sigma$} \\ \hline
 \end{tabular} %}
 \caption{Value ranges of neutrino masses, which are obtained from
  Eqs.~(\ref{Eq:IH}) and~(\ref{Eq:Val_DMij}). In addition to considering the
  mass constraint on heavier neutrino 
  $m_{ \nu_{3[2]} }^{2} \geqslant \Delta m^{2}_{31 [23] }$, and the relation
  $\sum m_{\nu_{i}} < 0.23$~\cite{Ade:2015xua}.}
 \label{Tab:masas_nu}
\end{table}
% 
%%%%%%%%%%%%%%%%%%%%%%%%%%%%%%%
\subsection{~{\it The likelihood test $\chi^{2}$} \hspace{12cm} \label{subsec:analysischi}} 
%Analysis of $\chi^{2}$ fit \label{subsec:analysischi}}
%%%%%%%%%%%%%%%%%%%%%%%%%%%%%%%
In order to verify the viability of our hypothesis of assert that all fermion 
mass matrices have the same generic shape, namely an four-zero texture, we make a 
likelihood test $ \chi^{2}$ in which the estimator function  is defined as:
%To test the viability of our hypothesis that assert all fermion mass matrices 
%have a universal shape, particularly a four-zero texture, we make a likelihood 
%analysis in which the estimator function $ \chi^{2}$ is defined as:
% 
\begin{equation}\label{eq:chicuadradadef}
 \chi^{2} = 
 \sum_{i<j}^{3} 
 \frac{ \left( 
  \sin^{2} \theta_{ij}^{ ^{\text{exp} } } 
  - \sin^{2} \theta_{ij}^{ ^{ \text{th} } } 
  \right)^2 
 }{
  \sigma_{ \theta_{ij} }^{2} 
 }.
\end{equation}
Here, the superscript $th$ states the theoretical expressions of mixing angles 
obtained from the Eqs.~(\ref{Eq:pmns:comp}) and~(\ref{Eq:senoscuadrados}), while 
the terms with superscript {\it exp} states the experimental data with uncertainty 
$\sigma_{\theta_{ij}}$. The experimental data for mixing angles considered in this analysis are given in
Table~\ref{Tab:tablasenos}~\cite{Forero:2014bxa}.
%Experimental data taken in this analysis are given in 
%Table~\ref{Tab:tablasenos}~\cite{Forero:2014bxa}.
%
\begin{figure}[t]
  \begin{tabular}{cc}
   \includegraphics[width=0.34\textwidth]{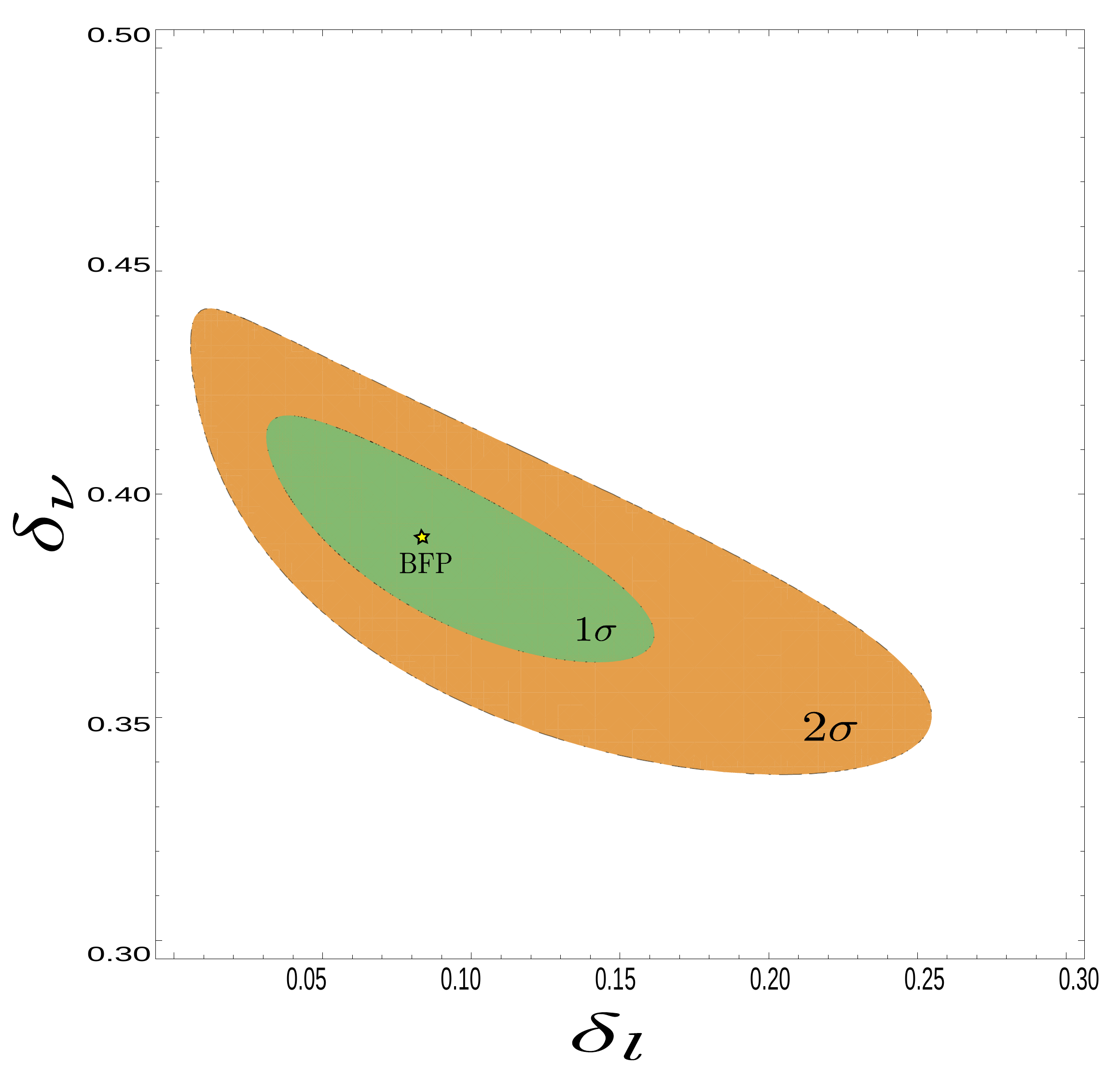} &
   \includegraphics[width=0.4\textwidth]{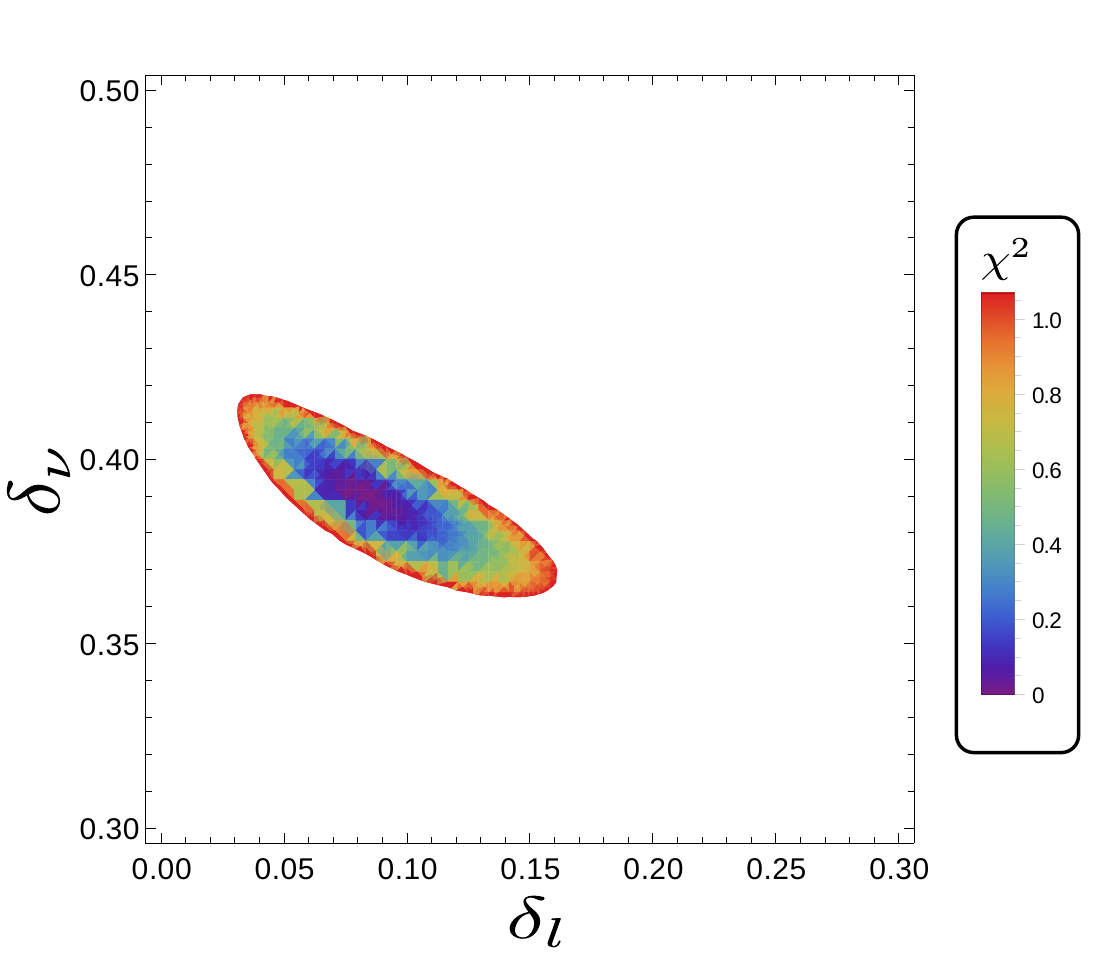}
  \end{tabular}
 \caption{For normal hierarchy. In the left graph, we show the allowed region of 
  the parameters $\delta_{l}$ and $\delta_{\nu}$. }
 \label{fig:m1JN}
\end{figure}

From expressions in 
Eqs.~(\ref{Eq:pmns:comp}),~(\ref{Eq:senoscuadrados}) and~(\ref{Eq:IH}), we can 
see that in general the $\chi^{2}$~function depends on five free parameters 
$\chi^{2} = 
\chi^{2} \left( \Phi_{1}, \Phi_{2}, \delta_{l}, \delta_{\nu}, m_{\nu_{3[2]}} 
\right)$. But with help of the analysis performed in the previous section, the 
heaviest neutrino mass is not considered like a free parameter because its 
numerical values are determined from the experimental data. Hence, the 
$\chi^{2}$~function has only four free parameters. 
\begin{table}[!htbp]
 \centering
 \begin{tabular}{lccc} \hline 
  {\footnotesize {\bf Parameter}} & 
   {\footnotesize {\bf BFP$\pm 1\sigma$}} & 
   {\footnotesize {\bf 2$\sigma$}} & 
   {\footnotesize {\bf $3\sigma$}} \\ \hline 
  {\footnotesize $\sin^{2} \theta_{12} (10^{-1})$} & 
   {\footnotesize $3.26\pm0.16$} & 
   {\footnotesize $2.92-3.57$} & 
   {\footnotesize $2.78-3.75$} \\ \hline  
  {\footnotesize $\sin^{2} \theta_{23} (10^{-1})$ [NH]} & 
   {\footnotesize $5.67_{-1.24}^{+0.32}$} & 
   {\footnotesize $4.14-6.23$} & 
   {\footnotesize $3.93-6.43$} \\ 
  {\footnotesize $\sin^{2} \theta_{23} (10^{-1})$ [IH]} & 
   {\footnotesize $5.73_{-0.39}^{+0.25}$} & 
   {\footnotesize $4.35-6.21$} & 
   {\footnotesize $4.03-6.40$}  \\ \hline
  {\footnotesize $\sin^{2}(\theta_{13})(10^{-2})$ [NH]} & 
   {\footnotesize $2.26\pm0.12$} & 
   {\footnotesize $2.02-5.20$} & 
   {\footnotesize $1.90-2.60$} \\ 
  {\footnotesize $\sin^{2}(\theta_{13})(10^{-2})$ [IH]} & 
   {\footnotesize $2.29\pm0.12$} & 
   {\footnotesize $2.05-2.52$} & 
   {\footnotesize $1.93-2.65$} \\ \hline 
 \end{tabular}
 \caption{Experimental results  of neutrino mixing angles in the ranges 
  $1\sigma,\, 2\sigma$ and $3\sigma$~\cite{Forero:2014bxa}. }
 \label{Tab:tablasenos}
\end{table}

Now to perform the likelihood test $\chi^2$, we consider that the neutrino masses, 
given in the Table~\ref{Tab:masas_nu}, run into the range of $2\sigma$. The 
values for lepton masses in MeV's are~\cite{Agashe:2014kda}
\begin{equation}
 \begin{array}{l}
  m_{e}=0.5109998928 \pm 0.000000011, \; 
  m_{\mu}=105.6583715\pm 0.0000035, 
  \; \textrm{and} \; 
  m_{\tau}=1776.82\pm 0.16.
 \end{array}
\end{equation}
Then, as result of the minimizing procedure of the $\chi^{2}$~function, for normal 
hierarchy in neutrino masses we obtain that the values of free parameters in the best 
fit point (BFP) are the following:
\begin{equation}\label{eq:bfp}
\begin{array}{ll}
 \Phi_1 = -6.789 \times10^{-1} \, \textrm{rad}, &
 \Phi_2 =  2.815 \, \textrm{rad}, \\
 \delta_l = 8.355\times10^{-2}, &
 \delta_\nu = 3.90\times10^{-1}, \\
 m_{\nu_{3}} = 5.00 \times10^{-2}  \,\textrm{eV}, & 
 \chi_{\min}^{2}=1.643\times 10^{-9}.
\end{array}
\end{equation}
%
%
%The neutrino masses minimization process is run into the range of $2\sigma$, the 
%obtained results are shown in Table~\ref{Tab:masas_nu}. Input parameters values 
%for lepton masses are $m_{e}=0.5109998928 \pm 0.000000011\, \textrm{MeV}$, 
%$m_{\mu}=105.6583715\pm 0.0000035\, \textrm{MeV}$ and $m_{\tau}=1776.82\pm 
%0.16\, \textrm{MeV}$~\cite{Agashe:2014kda}.
%
%In concordance with~(\ref{Eq:pmns:comp})~and~(\ref{Eq:senoscuadrados}), 
As mentioned above the $\chi^{2}$~function depends on four free parameters and three 
physical observables. Therefore, this function has minus one degrees of 
freedom, whereby we only can obtain the BFP.
%$\chi^{2} = \chi^{2} \left( \Phi_{1}, \Phi_{2}, \delta_{l}, \delta_{\nu} 
%\right)$. 
%However, this function depends just of three physical observables and has menus %one degrees of freedom, whereby we only can obtain the BFP. 
%which is computed  and the parameters values are the following:
%
%\begin{equation}
%\label{eq:bfp}
%\begin{array}{ll}
%\Phi_1=-6.789\times10^{-1} \, \textrm{rad}, &
%\Phi_2=2.815  \, \textrm{rad}, \\
%\delta_l =8.355\times10^{-2}, &
%\delta_\nu=3.90\times10^{-1}, \\
%m_{\nu_{3}}=5.00 \times10^{-2}  \, \textrm{eV}, & 
%\chi_{\min}^{2}=1.643\times 10^{-9}.
%\end{array}
%\end{equation}
%
However, from Eq.~(\ref{eq:bfp}) we know the numerical values for the free 
parameters in the BFP. So, a new analysis is performance fixing the CP violation 
phase, since this is the parameter less known from the experimental point of view. 
But, nowadays there are several experiments focussed on its measurement. Then, for a 
normal hierarchy in leptonic mass spectrum,  we fix the value of phases $\Phi_1$ and 
$\Phi_2$, as well as the heaviest neutrino mass $m_{\nu_{3[2]}}$ to the values given 
in Eq.~(\ref{eq:bfp}). 
So, the $\chi^{2}=\chi^{2}(\delta_l,\delta_\nu)$ function implies one degree of 
freedom.
%Last on chance us to obtain the allowed 
This last choice allows us to obtain the parameter regions 
%which $\chi^{2}$~function is enough small and phenomenological viable 
at different confidential levels. 
The results related to these regions are shown in Figure~\ref{fig:m1JN}.
%where one see the regions corresponding to $\chi^{2}(\delta_l,\delta_\nu)$ at $2\sigma$ or 90\% C.L. , respectively, BFP as well. That is, figure show the allowed region in parameter space ($\delta_l,\delta_{\nu}$) taken into account the BFP data. 

%These  BFP parameters values are taken to compute the neutrinos mass theoretical 
%ranges into the scheme of the two hierarchies, NH and IH. Furthermore, the %theoretical expressions of the mixing  angles squared sine are analyzed, and %finally, we compute the $\vpmns$.
%
%
\begin{figure}[h]
  \begin{tabular}{cc}
   \includegraphics[width=0.4\textwidth]{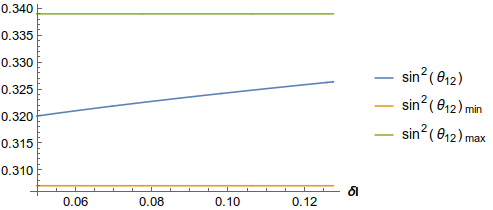} & 
   \includegraphics[width=0.4\textwidth]{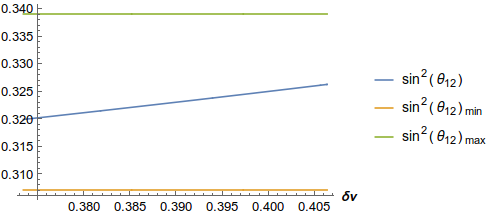} \\ 
   \includegraphics[width=0.4\textwidth]{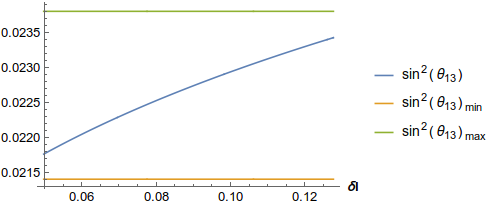} &
   \includegraphics[width=0.4\textwidth]{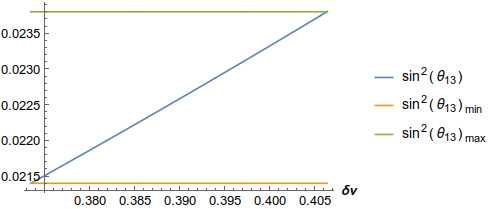} \\
   \includegraphics[width=0.4\textwidth]{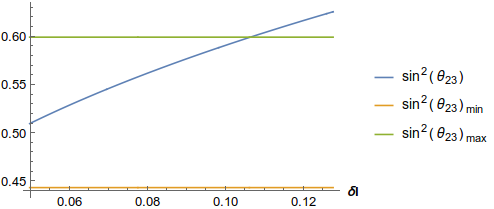} &
   \includegraphics[width=0.4\textwidth]{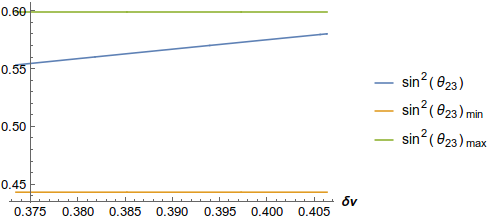}
  \end{tabular}
  \caption{For normal hierarchy. In the left graph, we show the allowed region of 
   the parameters $\delta_{l}$ and $\delta_{\nu}$ to $95\%$ C. L.. }
 \label{fig:sines}
\end{figure}
%

%%%%%%%%%%%%%%%%%%%%%%%%%%%%%%%%%%%%%%%
\subsection{ {\it The lepton mixing angles} \hspace{11.3cm} \label{subsec:analysisangles}}
%%%%%%%%%%%%%%%%%%%%%%%%%%%%%%%%%%%%%%%
Here, considering the results of the above likelihood test we study the sine of 
flavour mixing angles given by Eq.~(\ref{Eq:senoscuadrados}), as well as the PMNS 
matrix. 
In Figure~\ref{fig:sines}, we show the range of theoretical values obtained at 
$\pm1\sigma$ as the experimental edge values given in Table~\ref{Tab:tablasenos}. 
One can note that for both $\delta_l \& \sin (\theta_{12,13,23})$, and for 
the $\delta_{\nu} \& \sin (\theta_{12,13,23})$, results are inside the region of 
$1\sigma$.

As an immediate result of the above likelihood test~$\chi^{2}$, 
the flavour mixing matrix ${\bf V}_{\rm PMNS}$ is numerically computed, 
at $1\sigma$ C.L. 
\begin{equation}
 {\bf V}_{\rm PMNS}=
 %\resizebox{12cm}{!} {
 \begin{pmatrix}
  8.13 \times 10^{-1} \pm 6.06 \times 10^{-3} & 
   5.62 \times 10^{-1} \pm 6.38 \times 10^{-3} & 
   1.50 \times 10^{-1} \pm 9.16 \times 10^{-3} \\
  2.40 \times 10^{-1} \pm 3.91 \times 10^{-2} & 
   5.25 \times 10^{-1} \pm 4.23 \times 10^{-2} & 
   7.44 \times 10^{-1} \pm 4.21 \times 10^{-2} \\
  4.94 \times 10^{-1} \pm 2.68 \times 10^{-2} & 
   5.75 \times 10^{-1} \pm 3.36 \times 10^{-2} & 
   5.60 \times 10^{-1} \pm 5.81 \times 10^{-2} 
\end{pmatrix}.
\end{equation}

In the above section we have seen that in our theoretical framework, 
2HDM+$3\nu$, 
where the fermion mass matrix have a four-zero texture shape. We can reproduce 
the values of oscillation parameters in a very good agreement with the last 
experimental data. The next step in this study shall be to investigate the 
phenomenological implications of these results for the neutrinoless double beta 
decay ($0\nu \beta \beta $) and the CP violation in neutrino oscillations in 
matter.
%
%%%%%%%%%%%%%%%%%%%%%%%
\section{~Conclusions \hspace{13.6cm} \label{sec:conclusions}}
%%%%%%%%%%%%%%%%%%%%%%%
%
In the theoretical framework of Two Higgs Doublet Model type III  plus 
massive neutrinos (2HDM-III+$3\nu$), 
we shown that can be done we outlined a unified treatment for the 
fermion mass matrices in the theory. 
The active neutrinos are considered as Majorana particles and their 
masses are computed through the  type-I seesaw mechanism, 
%which implies an hybrid treatment of the neutrinos masses. By one side, 
where the right-handed neutrinos are introduced in the model as a 
%low 
singlet under the action of the gauge group of the Standard Model. 
%On the other hand, the active neutrinos are considering as Majorana particles. 
In such a treatment, the mass matrices of Dirac and right-handed neutrinos are 
represented with a four-zero texture ansatz, which implies that the mass 
matrix of left-handed neutrinos have also this shape with four-zero texture. 
In fact, all Dirac fermion mass matrices are represented with the same generic 
Hermitian matrix with four-zero texture and a normal hierarchy in the mass 
spectrum. 
%
%Accordingly, Majorana neutrinos mass matrix  $\mathbf{M}^R$  is a complex %symmetric %matrix.  
Theoretical expressions were derived for the elements of $\vpmns$ matrix in 
function of lepton masses, 
%$\widetilde{m}_{_{\nu_{i}}}$ $(i=1,2)$, 
two phases $\Phi_{1}$ and $ \Phi_{2}$ associated with the CP violation, and two 
parameters $\delta_{ \nu }$ and $\delta_{\nu}$ which are related with the Yukawa 
matrices of 2HDM-III.   
%$\vpmns(\widetilde{m}_{_{\nu_{i}}}, \eta_{1}, \eta_{2}, \delta_{l},\delta_{\nu})$.
From the theoretical relations of the differences of the squared neutrino masses,  
and the experimental results reported by the Planck Collaboration and neutrino 
oscillation experiments, we obtain the allowed values for the neutrino masses. 
%
%From the theoretical differences squared mass relations $\Delta^{2} m_{ij}$ 
%and experimental results reported by Planck Collaboration and neutrino %oscillations experiments, we obtain the allowed values for the neutrino masses. 
%associated with the light and heavier neutrinos masses for each hierarchy 
%(NH and IH). This allows the mixing matrix $\vpmns$ only depends on the heaviest neutrino mass. 
The parameter space exploration is done by means of likelihood test $\chi^{2}$;
%function that allows us to measure the difference between theory and experimental statistics. In turn, 
this allowed us to find the allowed regions of the parameters $\delta_{\nu}$ and
$\delta_{l}$ at  70\% and 95\% C.L.  for a normal hierarchy, as well as, the best 
fit point (BFP), and the mixing matrix $ \vpmns$  at  70\% C.L. 
Finally, it is observed that the mixing angle as function of $\delta_{\nu}$ and $\delta_{l}$ are in very good agreement with experimental data.

%%%%%%%%%%%%%%%%%%%%
%%	A C K N O W L E D G M E N T S
%%%%%%%%%%%%%%%%%%%%
\section{Acknowledgments \hspace{13cm} }
%\begin{acknowledgments}
This work has been partially supported  by \textit{CONACYT-SNI (Mexico)}. 
ERJ acknowledges the financial support received from \textit{PROFOCIE (Mexico)}. 
F.G.C. acknowledges the financial support received from Mexican grants 
\textit{CONACYT} 236394, 132059, and PAPIIT IN111115.
%\end{acknowledgments}
%%%%%%%%%%%%%%%%%%%
%	BIBLIOGRAPHY
%%%%%%%%%%%%%%%%%%%
% References
%\nocite{*}
%\bibliographystyle{aipnum-cp}%
%\bibliography{biblioteca}%
%
%
%
\end{document}